\documentclass{PoS}

\title{After the SKA - Radio Astronomy in 2049}

\ShortTitle{After the SKA - Radio Astronomy in 2049}

\author{\speaker{Lisa Harvey-Smith}\\
              CSIRO Astronomy \& Space Science\\
        E-mail: \email{lisa.harvey-smith@csiro.au}}

\abstract{The concept of a Square Kilometre Array was developed to ensure that progress in Radio Astronomy in the early 21st Century continued at the same impressive pace as was achieved during the first 50 years. The SKA telescope is designed to pave that road to greater and greater sensitivity. So what technical challenges does the project face and what key innovations will drive the success of the SKA? What will the next Radio Astronomy mega-science project look like? In this article the author discusses the likely avenues of progress in the coming decades and comments on the status of radio astronomy in 2049 - the author's 70th (and presumably her retirement) year.}

\FullConference{Resolving The Sky - Radio Interferometry: Past, Present and Future \\
		 April 17-20,2012\\
		 Manchester, UK}

\begin{document}

\textit{There is only one perfect road and that road is ahead of you, always ahead of you - Sri Chinmoy} \\

\section{Introduction}
Although predictions in science often turn out to be wrong, I decided to write this article for two reasons. First, scientists should be entirely comfortable with being wrong. Second, I believe that learning from the history of science is valuable in that it provides perspective and clues on how best to proceed. I hope that this article will encourage discussions on global approaches to radio astronomy during the detailed design of the Square Kilometre Array (SKA) and beyond. 

Professor Ron Ekers elucidated perfectly the need for an SKA radio telescope when he said ``\textit{If the improvement in sensitivity has reached a ceiling, the rates of new discoveries will decline and the field of radio astronomy will become uninteresting and die out. On the other hand if we can shift to new technology or find new ways to organise our resources, the exponential increase in sensitivity can continue}".

\begin{figure}[htbp]
\begin{center}
\includegraphics[width=430pt]{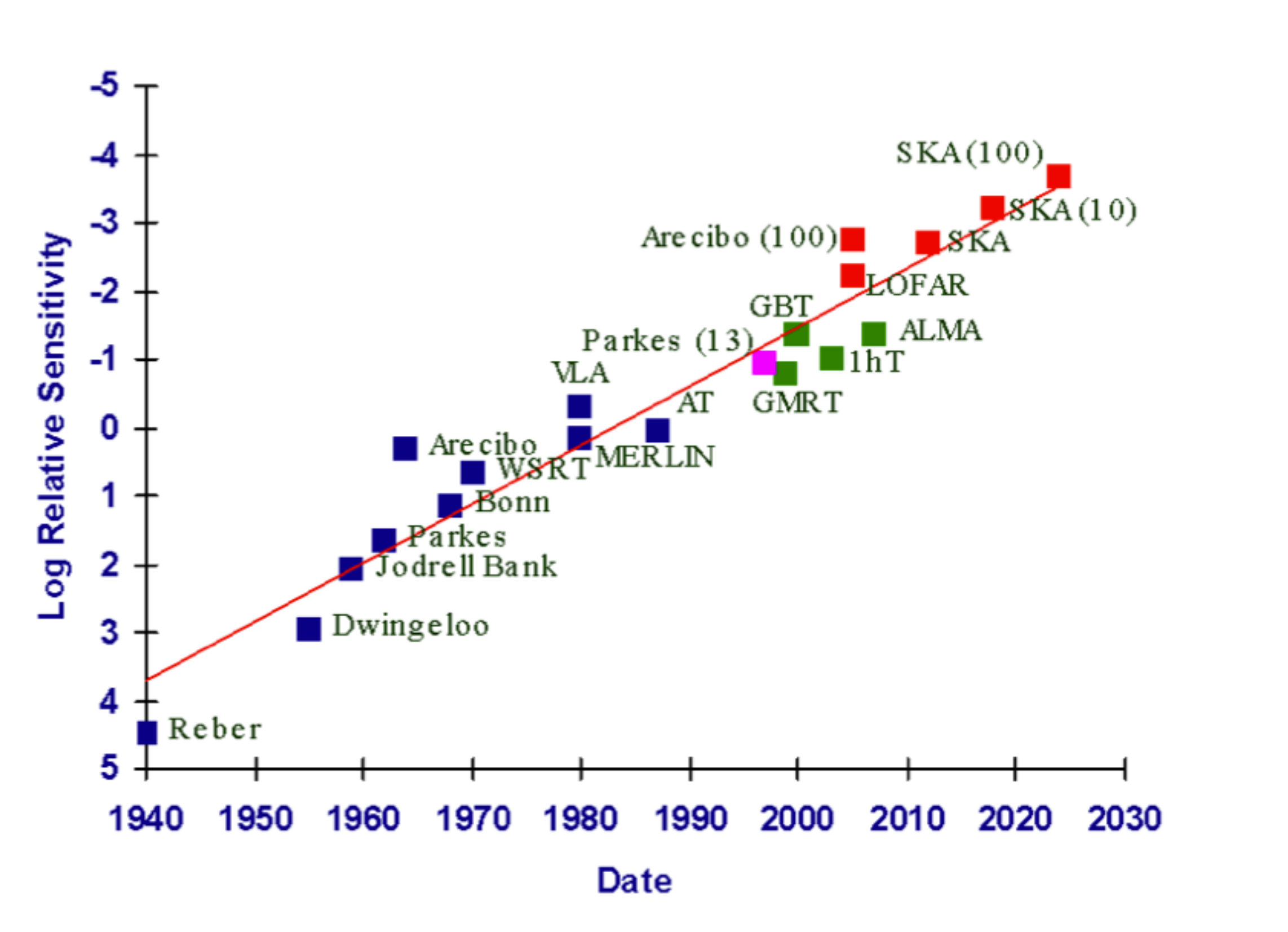}
\caption{Plot reproduced from Ekers \& Bell (2001), showing the historical and projected sensitivity of radio telescopes assesed at the turn of the 21st Century}
\label{fig1}
\end{center}
\end{figure}

The historical and projected improvement in the sensitivity of radio telescopes at the turn of the 21st Century is shown in Figure \ref{fig1} from \cite{ekers}. We can hope to stay ahead of this curve in three ways: \\
(1) {\bf Innovate} with smart receivers, intelligent software, clever array design, advanced RFI monitoring and mitigation techniques 
\\
(2) {\bf Saturate} using larger collecting area, wide bandwidth, larger field-of-view, high time resolution, powerful computing and 
\\
(3) {\bf Co-operate}, through broader global science collaborations, leveraging existing resources and utilising citizen science. 

The greatest challenge will be to realise all of this within an achievable budget. Instead of trying to include every technological possibility within SKA, it would be wise for astronomers to also plan a long-term staged approach to the realisation of a highly sensitive, global connected radio telescope beyond the lifetime of the SKA project. 

\section{Imminent Innovations in Radio Astronomy}

\subsection{Phased Array Receivers}

As the radio astronomy receiver of choice for the next few decades, the march of focal plane arrays or phased arrays\footnote{The terms focal plane array and phased array are used interchangeably in this article and imply the inclusion of all types of beamformed arrays of radio receivers.} looks unstoppable. The major benefits of phased arrays include high survey speeds, wide instantaneous fields-of-view and the ability to quickly steer several independent beams on the sky with no moving parts. As such, focal plane arrays are ideal for rapid sky surveys of neutral hydrogen and radio continuum emission. Coupled with a good instantaneous point spread function, this also enables the detection and study of transient and variable radio sources.

Technical innovation is what underpins the current `freestyle' developmental stage of the international SKA project.  Recognising the potential of phased arrays, several of the SKA pathfinders and precursors, including the Murchison Widefield Array (MWA), The Australian SKA Pathfinder (ASKAP), the Precision Array for Probing the Epoch of Reionisation (PAPER), the Low-Frequency Array (LOFAR), the Long Wavelength Array (LWA), Aperture Tile in Focus (APERTIF) and Electronic Multi-Beam Radio Astronomy Concept (EMBRACE) employ this technology.  

Although the hardware is relatively simple, the performance of electronic components largely drives the performance of phased arrays. There are ongoing challenges for these cutting-edge instruments such as calibration and imaging dynamic range, to name just two. Nonetheless, continued improvements in low-noise amplifiers, digital signal processing, high performance computers and astronomical processing algorithms provide a clear upgrade path for improved performance in phased arrays for radio astronomy that will span decades.

\subsection{Advances in Computing: Moore's Law no more?}

The sheer scale of the SKA project presents many computing challenges, due to the implementation of broad-band, high spectral resolution, high time resolution and wide-field instrumentation all at once. Processing the resulting data, storing them and carrying out science data processing will probably provide the most serious constraints on the scientific utility of the SKA in the short term.

For several decades the number of transistors that can be placed on an integrated circuit has doubled approximately every two years - the effect described as Moore's Law \cite{moore}. Some believe that Moore's Law is a self-fulfilling prophesy, entrenched in human ingenuity. But most commentators agree that there are some physical limitations, for example transistor size, density of components, chip size and the limits of optical lithography in circuit fabrication.

For many years there has been talk of the demise of Moore's Law (e.g. \cite{kish},\cite{schaller}), as electronic devices approach what are presumed to be fundamental limits. Recent work, including the demonstration of a single-atom transistor \cite{fuechsle} could signal the physical limits of Moore's Law for small silicon-based circuits. Work-arounds such as increased chip size or multiple layered chips may provide some respite.   

In terms of processing power, Quantum Computing could in the longer term take over the mantle. Recent results \cite{lopez} describe a quantum computer that has successfully factorised the number 21. More challenging problems have been tackled by a diamond based quantum computer created by a team led by the University of Delft \cite{vds}. The current commercial leader in the field is the room-sized ``Quantum Optimiser", based on a cryogenically cooled 128-qubit superconducting processor that was sold by the Canadian company D-Wave to Lockheed Martin for US \$10 million. Skeptics notwithstanding, the field has enormous potential in the field of encryption/decryption and for that reason one would expect a well-funded and sustained research and development effort over the coming decades. By 2049, Quantum computers could possibly have superseded classical computers, which could prove vital in the upgrade path for SKA from lower to higher bandwidth, field of view, spectral and time resolution and high angular resolution all-sky surveys.

\section{Powering the SKA}

The ongoing power requirement for a single site implementation of SKA has been estimated\footnote{SKA Site Advisory Committee Report: www.skatelescope.org/.../40391\_120216\_SSAC.Report\_web.pdf} at 110 MW. This is required to drive the thousands of dishes, for cooling, beamforming, correlation, data transport and science data processing. In order to unlock the capabilities of new generation radio receivers such as phased arrays and ultra-wide bandwidth feeds, it will be necessary to take every available step towards reducing the power requirements of these systems. Only then can we hope to unlock the full scientific potential of the SKA.

\subsection{Low-power quantum electronics}

Reducing power consumption will reduce costs and thereby increase the science capabilities of the SKA. Using current technology, it has been estimated that SKA will employ approximately 10$^{16}$ transistors.\footnote{John Bunton, private communication}  Current microchips use Field Effect Transistors (FETs), which have a potential difference of approximately 1V. Development work is underway on a new generation of microchips, which will combine conventional and tunnel-FET technology. Current prototypes run on only 0.2V, which can reduce the power consumption by up to 100 times \cite{ionescu}. 

According to Professor Adrian Ionescu in a press statement\footnote{Physics World: http://phys.org/news/2011-11-impending-revolution-low-power-quantum.html} issued by the Guardian Angels project (a European collaboration on low-energy technology development), prototype microchips using quantum-FET hybrid transistors have been made by IBM in Zurich and the CEA-Leti in France. Mass production of these devices is expected by around 2017. If low-power devices are available on the timescale of SKA, this may provide considerable savings in energy usage. Of course, the project must consider the trade-off between the possibly additional cost of these components relative to older models versus the on-going operational cost savings due to their reduced power requirements. 

\subsection{Renewable Energy for Radio Astronomy}

As the performance of the SKA may be power limited \cite{skamemo84}, it will make sense to generate a significant proportion from renewable sources, thus insulating the project from market price fluctuations. Radio Astronomy could potentially benefit via access to new funding streams, and exemption from pollution tariffs. As a publicly-funded global scientific project, the SKA should aim to be a model of good practice to be adopted by other observatories and scientific institutions. 

The Australian SKA Pathfinder is already prototyping a number of renewable energy schemes. The on-site data processors for ASKAP will be cooled by a geothermal system serviced by a shallow bore-field, which was completed in 2012. Depending on the eventual design of the SKA core, this renewable cooling system may be scaled to meet the on-site data processor cooling requirements for SKA at the Murchison Radio-astronomy Observatory. 

ASKAP's science data processor, housed at the Pawsey Supercomputing Centre for SKA Science is the subject of a multi-faceted geothermal energy project. The Commonwealth Scientific and Industrial Research Organisation (CSIRO) Geothermal Project is a \$19.8 million renewable energy initiative. It incorporates a geothermal deep well project, which is investigating the viability of geothermal power generation from a hot sedimentary aquifer 3.3~km below the city of Perth in Western Australia. The scheme also has a production well, dug to a depth of 2.2~km to enable cooling solutions to be implemented at the supercomputer. 

A related project has funded studies on a shallow (140~m depth) aquifer heat rejection system for the Pawsey Supercomputing Centre. The bore field drilling will progress to the next stage in 2013 when the Pawsey building has been completed. Finally, the Sustainable Energy for SKA project will provide off-the-grid renewable energy generation, with enhanced solar penetration for the Murchison Radio-astronomy Observatory. On the back of such developments, it seems reasonable that SKA Phase 2 aims for renewable energy providing at least 50\% of the project's power and cooling needs.

\section{Broadening Global Collaborations}
\subsection{Science in the Pub: TAVERN - a global transient research network}

Wide field-of-view SKA Pathfinders will transform time-domain astronomy, leading to the discovery of new classes of object and unexplained phenomena. These will require detailed multi-wavelength study at reasonably high angular resolution (to allow cross-identification) and in some cases extended monitoring in order to gather useful information about the origin and nature of such events. 

Some of the scientific potential in this field can be squeezed from the data in post-processing, for example through mining of virtual observatory databases. However, for much of the science a rapid follow-up, or better still a co-ordinated approach to the resolution, cadence and sky coverage of surveys at different frequencies would be beneficial.

In order to make a meaningful impact on this emerging field of astrophysics, it would be prudent for the astronomical community to start planning a formal co-ordinated effort between the new generation of wide field-of-view telescopes in order to exploit this mine of information on transient and variable sources. This co-ordination framework should inform the design and operational plans for such instruments. If such an approach is not taken, it may be that the SKA and other wide-field instruments simply select the `low hanging fruit' in the time domain, rather than truly exploiting their full potential. One might think of this Transient and Variables Extraterrestrial Research Network as a sort of `public house' for astronomical research - a place to meet, share data and drink in the secrets of the time-domain universe.

\subsection{Citizen Science}

Citizen Science is scientific research carried out by non-specialists on a voluntary basis. Some of the most popular Citizen Science projects in the field of astronomy use idle home computers to process data (e.g. SETI@Home\footnote{\texttt{http://setiathome.berkeley.edu/}}) and employ the visual and cognitive aptitude of citizen scientists to classify complex objects (e.g. Moon Zoo, Galaxy Zoo, The Milky Way Project)\footnote{\texttt{http://www.zooniverse.org/}}. There are many such projects around the world and they are becoming increasingly well-organised and successful.

As we enter a particularly data-rich era of radio astronomy, there is considerable potential for citizen science to make a meaningful contribution. However, the ability for citizens science projects to exploit the results of SKA observations will rely on the open-source nature of the data. 

%That is not to say contributing countries within the SKA project cannot insist on proprietary periods for data. This would not in any way negate the possibility of citizen science projects as radio astronomy is likely to be data-rich and relatively resource poor for some decades. The number of professional radio astronomers in the world is not enough to immediately write all possible science papers on-the-fly as data comes out of the telescope. In fact, citizen science within the SKA could and should become extremely well organised, in collaboration with professional astronomers, to allow the publication of results. 

A potential stumbling block is the freedom of information transfer between nations. The prospect of controlling content on the world wide web has recently been discussed within several jurisdictions, for example mandating internet service providers to block their users from accessing certain websites due to intellectual property infringements. Governments in several countries already restrict access to certain domains, citing issues of national security. Hopefully this will not become an issue for the free exchange of scientific data, but the community must remain vigilant to such proposals in order that controls do not adversely affect the flow of scientific data.

\subsection{Towards global VLBI standards}

There  are many compelling science cases for high sensitivity observations using the SKA with very long baselines \cite{godfrey}. This will be provided by the SKA project itself but may also be augmented by joint observations between the SKA and existing radio telescopes. In Australia, CSIRO and The University of Tasmania have operated the Long Baseline Array as a VLBI National Facility since 1997 (with correlation now carried out by Curtin University of Technology). The network also uses the Warkworth 12~m antenna in New Zealand as a regular part of the array. Existing and planned telescopes in Asia-Pacific are ideally placed for even longer baselines to ASKAP and the SKA telescopes in Australasia. South Africa is a member of the European VLBI Network and has also signalled its ambition to develop a trans-African VLBI network, using decommissioned telecommunications antennas across the continent.

As global VLBI networks expand, new antennas join existing networks and telescopes receive upgrades to use more modern disk-based recording and real-time e-VLBI. As a consequence, a number of new and sometimes incompatible data standards have appeared around the world. The lack of standardisation in VLBI data has led to considerable wasted effort in converting these data before correlation becomes possible. In the early 2000's the VSI-E specification \cite{lapsley} was implemented amongst some members of the VLBI community but was never adopted universally. More recently, The VLBI Data Interface Specification (VDIF) \cite{whitney} provides a standard interchange format for VLBI data, although it does not include a data-transport protocol.

Clearly, with an \textit{ad hoc} array using a small number of telescopes, the processing overhead of data format conversion is a minor inconvenience. But in the era of SKA pathfinders and eventually several hundreds or thousands of dishes, having an agreed standard, agreed through the International Astronomical Union, would be beneficial.

\section{Upgrade Paths and Future Instruments}

By their very nature, large number - small diameter arrays such as the SKA can be upgraded and augmented as far as the existing design and the physical environment allows. The best way to see further into the universe is to add collecting area, as no post-processing can add to the sensitivity of a hydrogen-detecting telescope beyond this fundamental limit. 

In this way, the limit to SKA's upgrade path is suitable landmass, serviced by optical fibre and power and free from radio frequency interference. One might suppose that a square kilometre array can readily become a two square kilometre array and so on provided the build is practical and affordable. Herein lies the problem - spectrum usage is increasing rapidly and radio-quiet areas are becoming increasingly rare. 

If interference becomes a fundamental limitation then leaving Earth would seem to be the only option. The environmental and monetary cost of placing millions of tonnes of metal into space is currently prohibitive. Radio telescopes on the far side of the moon have been suggested, but I do not expect to see a scientific radio telescope on the moon before 2049 unless a dramatically cleaner, cheaper and safer means of transporting large payloads is found. 

The solution to bypassing the increased spectrum usage on earth may be larger collecting area, higher time resolution, robust digital sampling, smart adaptive-beam instruments and possibly for some observations \emph{narrower bandwidth receivers}. Knowing where and when strong sources of RFI are likely to appear and avoiding them may be our best defence. Provided our instantaneous sensitivity is very good in the time-frequency parameter space that is unaffected by strong interference, we can still carry out very sensitive radio astronomical observations. We may simply have to rely less on gaining sensitivity by averaging over large contiguous frequency ranges and long contiguous time periods. There is clearly much work to be done on developing best practice in RFI monitoring and mitigation as part of the detailed design of the SKA.

\section*{Summary}
The Square Kilometre Array is a robust concept with considerable upgrade potential, provided this path is carefully woven into the design from the outset. As scoped, the SKA should still be carrying out world-leading science in 2049. Following the exponential sensitivity growth of radio telescopes (Figure~\ref{fig1}) and Moore's Law, additional collecting area coupled with increasingly sensitive receivers and powerful processors may prove the best upgrade path of all. 

\section*{Acknowledgements}

\noindent Many colleagues, including Jan Noordam, Robert Braun, Ron Ekers, Michelle Storey, John Bunton, Chris Phillips, Erik Lensson, Carol Wilson and Tony Mulry provided valuable insight.

\end{document}